%% file: template.tex
\pdfoutput=1

\RequirePackage{fix-cm}
\documentclass[smallextended]{svjour3}       % onecolumn (second format)
\smartqed  % flush right qed marks, e.g. at end of proof
\usepackage{graphicx}
\usepackage{algorithm,algpseudocode}
\usepackage{xurl}
\usepackage{amsmath}
\usepackage{amssymb}
\usepackage{makecell}
\usepackage{rotating}
\usepackage{multirow}
\usepackage{dirtytalk}
\usepackage{placeins}
\begin{document}

%\title{Learning from Irregularly Sampled Data with Nadaraya-Kernel Regression Networks (NWNet) for Endomicroscopy Super-resolution}
\title{Learning from Irregularly Sampled Data for Endomicroscopy Super-resolution: A Comparative Study of Sparse and Dense Approaches}
\titlerunning{Learning from Irregularly Sampled Data for Endomicroscopy Super-resolution}

\author{
       Agnieszka Barbara Szczotka$^1$
       \and
       Dzhoshkun Ismail Shakir$^2$
       \and
       Daniele Rav\`{\i}$^3$
       \and
       Matthew J. Clarkson$^1$
       \and 
       Stephen P. Pereira$^4$
       \and
       Tom Vercauteren$^2$
       }
\institute{
1.  Wellcome / EPSRC Centre for Interventional and Surgical Sciences, University College London,
\email{agnieszka.szczotka.15@ucl.ac.uk}
\and
2. Department of Surgical and Intervention Engineering, King's College London
\and
3. Centre for Medical Image Computing (CMIC), Department of Computer Science, University College London
\and
4. UCL Institute for Liver and Digestive Health, University College London
}
\authorrunning{A.B. Szczotka et al.}
% F} % if too long for running head

\date{Received: 28 Nov 2019 / Accepted: 28 Nov 2019}

\maketitle
\sloppy
\begin{abstract}
\input{0abstract}
 %4-6 
\end{abstract}

\section{Introduction}
\input{1introduction}
\section{Related work}
\input{2related_work}

\section{Materials and methods}
\input{3data_intro}
\input{4data_simulation}
\input{5NW}
\input{6nw_implementation}

\section{Results}
\input{7results_validation}

\section{Discussion and conclusions}
\input{8conclusions}

\section*{Acknowledgement}
\input{9acknowladgment}
\section{Compliance with ethical standards}
\input{9conflict}

%\begin{acknowledgements}
%\end{acknowledgements}


\bibliographystyle{spmpsci}    
%\bibliographystyle{spbasic}
% mathematics and physical sciences
\bibliography{nw}   % name your BibTeX data base


\end{document}

%% file: 0abstract.tex
\label{abstract}
\textbf{Purpose:} Probe-based Confocal Laser Endomicroscopy (pCLE) enables performing an optical biopsy, providing real-time microscopic images, via a probe. pCLE probes consist of multiple optical fibres arranged in a bundle, which taken together generate signals in an irregularly sampled pattern. Current pCLE reconstruction is based on interpolating irregular signals onto an over-sampled Cartesian grid, using a naive linear interpolation. It was shown that Convolutional Neural Networks (CNNs) could improve pCLE image quality. 
Although classical CNNs were applied to pCLE, input data were limited to reconstructed images in contrast to irregular data produced by pCLE.
\textbf{Methods:} We compare pCLE reconstruction and super-resolution (SR) methods taking irregularly sampled or reconstructed pCLE images as input. We also propose to embed a Nadaraya-Watson (NW) kernel regression into the CNN framework as a novel trainable CNN layer. Using the NW layer and exemplar-based super-resolution, we design an NWNetSR architecture that allows for reconstructing high-quality pCLE images directly from the irregularly sampled input data. We created synthetic sparse pCLE images to evaluate our methodology. %We also use real pCLE data to test the applicability of models trained on synthetic data to original pCLE images.
%Our evaluation exploits a database of 181 images from 30 patients. 
\textbf{Results:} The results were validated through an image quality assessment based on a combination of the following metrics: Peak signal-to-noise ratio, the Structural Similarity Index.
% Our analysis indicates that the proposed solution 
%enables a principled adaptation of CNNs for processing sparse data. 
\textbf{Conclusion:} Both dense and sparse CNNs outperform the reconstruction method currently used in the clinic. The main contributions of our study are a comparison of sparse and dense approach in pCLE image reconstruction, implementing
trainable generalised NW kernel regression, and adaptation of synthetic data for training pCLE SR. 
% 242 words with latex and comments
% 150-250 words
%. In contrast, the proposed method produces compelling results

%% file: 1introduction.tex
\label{Intro}
\raggedbottom
Probe-based Confocal Laser Endomicroscopy (pCLE) is a recent optical fibre bundle based  medical imaging modality with utility in a range of clinical indications and organ systems, including gastrointestinal, urological and respiratory tracts~\cite{Fugazza2016}.\par
The pCLE probe relies on a coherent fibre bundle comprising many~($>$10k) cores that %: 1) have variable size and shape; 2)
are irregularly distributed across the field of view (FoV).
%3) exhibit variable light transmission properties, including coupling efficiency and inter-core coupling. 
The nature of image acquisition through coherent fibre bundles constitutes a source of inherent limitations in pCLE, having a direct, negative impact on the image quality. The raw data that the pCLE devices produce, therefore remain challenging to use for both clinicians and computerised decision support systems.\par
Raw pCLE images are distorted by a few artefacts such as a honeycomb pattern, and so need to be corrected before reconstruction. During calibration and restoration, the raw image is transformed into a vector of corrected fibre signals, and their locations in the space of the fibre FoV~\cite{le2004towards}. The irregular sampling domain of the signals can be accurately discretised as a set of locations in an over-sampled regular grid and then interpolated.
\par
Existing pCLE image reconstruction approaches typically use Delaunay triangulation to linearly interpolate irregularly sampled signals onto a Cartesian grid~\cite{vercauteren2006robust}. These interpolation methods allow to reconstruct the Cartesian image, yet do not enhance image quality nor take into account any prior knowledge of the image space except for regularisation related properties. Moreover, they are themselves prone to generating artefacts, such as triangle edge highlights or additional blur~\cite{vercauteren2006robust}.

%Incorporating prior information about pCLE images should reduce the uncertainty introduced by the reconstruction process and enable higher quality reconstructions.
% reconstruction improved by SR and combined
Reconstructed pCLE images can be post-processed by restoration and super-resolution (SR) techniques to improve image quality. It was shown that state-of-the-art CNN-based single-image super-resolution (SISR) techniques improve the quality of pCLE images~\cite{ravi2018effective}. A potential limitation in the current CNN approaches is that the analysis starts from already reconstructed pCLE images, including reconstruction artefacts as mentioned above.
\par
% which might hinder the full potential of deep learning (DL)
There are a few research focusing on allowing sparse data as CNN input~\cite{eldesokey2018propagating,hua2018normalized,uhrig2017sparsity}.
Generalising the conclusion from these studies to pCLE hints to the intuition that applying CNNs directly to irregularly sampled pCLE data is far from trivial.
We propose a solution that facilitates using sparse images as the input of the SR CNN directly, without the need for prior reconstruction, and also eliminating edge artefacts from input images and compare it to the classical SR methods and reconstruction algorithm.  
%Existing CNNs rely on shift-invariant convolution to unlock the potential of artificial neural networks for
%image processing and recognition tasks. When applied directly to fibre-bundle-modulated images, whose structure is highly non-shift-invariant, the shift-invariance of CNNs has a detrimental effect.
%Also, when dealing with small images, such as those produced with fibre bundles, the cropping effect of standard convolutional layers limits the usable depth of the CNNs. 
% solution

%The vast majority of DL techniques for image data relies on regularly sampled (Cartesian) images. Hence there is an unmet need for a unified, computationally-efficient, image
%reconstruction methodology that compensates for a range of limitations, including irregular sampling pCLE artefacts.
Since the vast majority of DL techniques, including SISR used for the pCLE reconstruction, rely on Cartesian images, we propose a unified, computationally-efficient,
methodology that generalises NW kernel regression as a part of DL framework and allows it to be optimised. %CNNs for irregularly sampled images and compensates for a range of limitations in pCLE reconstruction task.
The main focus of this work is to compare pCLE image reconstructions obtained from the classical interpolation method and dedicated DL architectures, applied directly to the irregularly sampled or reconstructed Cartesian images respectively.
\par
For the comparison study, we design a novel trainable convolutional layer called an NW layer, which integrates Nadaraya-Watson (NW) kernel regression~\cite{nadaraya1964estimating} into the DL framework, allowing effective handling of irregularly sampled data in the CNN network. It is the first work proposing a unique CNN architecture (referred to as NWNetSR), which takes advantage of the potential of both the NW kernel regression and the SISR to benefit pCLE image reconstruction.
To the best of our knowledge, we are the first to propose using NW kernel regression embedded in a CNN framework to design a network for medical image super-resolution reconstruction from irregularly sampled pCLE signals.
%We exploit the irregularly sampled pCLE data for image reconstruction tasks. 
%The rest of the paper is organised as follows. Section~\ref{related_work} gives a quick overview of the current state of the art methods which enable sparse image data to be used as an input for CNN. 
%Section~\ref{NW_section} presents the details of the NW layer and NWNet networks with the implementation details, used datasets and the training strategy.
%Section~\ref{results} presents quantitative image quality assessment (IQA) for evaluating the performance of NWNets in the context of image reconstruction. Section~\ref{conclusions} summarises the contribution of this research to pCLE imaging and deep learning research.

%. Echoing the impact of deep learning for classical images, NWNets will lead to a step change in performance on irregularly sampled image computing tasks.
%End-to-end deep learning is presently pushing the boundaries of medical image computing. s will take end-to-end training one step further by allowing ConvNet-like approaches on irregular sampling grids.

%It also increases input for computation which leads to lower time performance.It should lead to a better allocation of computational resources. 

%% file: 2related_work.tex
\label{related_work}
\paragraph{\textbf{pCLE image reconstruction:}}
\par
The pCLE image reconstruction  algorithm in current clinical use is based on Delaunay triangulation. It yields sharp images, but contains triangulation artefacts~\cite{tom_phd}.
\par
Image reconstruction from sparse signals has been widely studied.
Specifically, in the context of pCLE, Vercauteren et al. implemented reconstruction from scattered pCLE data with NW kernel regression using handcrafted Gaussian weighting kernels~\cite{vercauteren2006robust}. They demonstrated that the method efficiently reconstructs pCLE images and mosaics, at the price of some additional blur in comparison to Delaunay reconstruction.
\paragraph{\textbf{pCLE super-resolution:}} 
It has been shown that CNNs could effectively improve pCLE quality. In the study~\cite{ravi2018effective}, researchers compared the performance of state-of-the-art SISR networks in reconstructing SR pCLE images. Due to the lack of ground truth high-resolution (HR) images,  they used mosaicing to estimate synthetic HR images
% by cropping high-quality mosaics %via inversion of the mosaicing transformation, 
and simulated pCLE signal loss to create synthetic low-resolution (LR) images. They confirmed both quantitatively and qualitatively that CNNs trained on the synthetic data can improve the pCLE image quality.\par 
Another work showing a promising proof-of-concept for SR reconstruction uses DenseNet trained on pairs of LR and HR patches~\cite{izadi2018can}. The LR endomiscroscopy was simulated by bi-cubic down-sampling of HR confocal laser endomicroscopy (CLE). These CLE images are generated by the Pentax EC-3870FK, and they are not affected by same distortions as pCLE are, because the Pentax device does not use a fibre bundle as an imaging guide. They prove that their DL solution outperforms classical interpolation by recovering HR details and reduces pixelation artefacts when used to super-resolve synthetic LR images.\par
In the absence of HR pCLE images, also unsupervised blind super-resolution has been proposed. The work contributes a novel architecture based on adversarial training with cycle consistency for pCLE~\cite{ravi2019adversarial}. However, all these works use reconstructed Cartesian pCLE and CLE images as input to the networks.

\paragraph{\textbf{Sparse CNN inputs:}} While convolution layers are widely used, they have been identified as sub-optimal for dealing with sparse data~\cite{uhrig2017sparsity}. 
%Several approaches have been proposed to handle sparse data as input to CNN networks.
Much of the available literature on exploring sparsity in the context of CNN input deals with the irregular data in an intuitive but ad-hoc way:
non-informative pixels are assigned zero, creating an artificial Cartesian image.
%to be input to the network.
For example, Li et al. used that technique and assigned zeros to the missing points on an LR image~\cite{li2016vehicle}. A similar workaround is to use an additional channel to encode the validity of each pixel like in Kohler et al. they passed a binary mask to the network~\cite{kohler2014mask}.
These solutions suffer from the redundancy in image representation due to spurious data is being fed to the convolutional layers.%, Furthermore, this may lower the performance of the model optimisation.
\par
In a recent study, Uhrig et al. proposed a convolutional layer which jointly processes sparse images and sparse masks to achieve sparsity invariant CNNs~\cite{uhrig2017sparsity}. Their sparse layer is designed to account for missing data during the convolution operation by modelling the location of data points with the use of a mask. This is achieved by convolving the mask with a constant kernel of ones while optimising the solution through convolving the sparse image with trainable kernels.\par
Following the success of sparse CNNs,  Hua, J. et al. proposed to implement normalised convolution, as an extension of sparse convolution~\cite{hua2018normalized}. They showed that using shared positive kernels for convolution with both an image and a mask is beneficial for upsampling depth maps.  In both aftermentioned works, the information on sparsity is propagated to consecutive layers by the binary mask.\par
A demonstrated improvement of the proposed solutions is to use soft certainty maps, rather than propagating binary masks~\cite{eldesokey2018propagating}. These maps are produced by updating the mask with the convolution. This method worked well in a guided depth upsampling task and uses both RGB data and LiDAR to reconstruct depth maps.\par

%The critical difference between their approach and ours is that the NW layer performs a methodologically founded Nadaraya–Watson kernel regression, while the sparse CNN layer does not perform a typical regression, but instead uses a non-shared constant kernel for the convolution of the mask. 

%% file: 3data_intro.tex
\label{data_sim}
\par
Since common Image Quality Assessment (IQA) relies on  ground-truth images  used as a reference in metrics such as the Peak signal-to-noise ratio (PSNR), the lack of ground-truth high-resolution pCLE images makes it difficult to evaluate the quality of SR reconstructions.
\par
To address the lack of the HR pCLE, the authors in~\cite{ravi2017deep} proposed to use a first-generation SR method - an offline mosaicing - %based on registration,
to simulate HR endomicroscopy. They used mosaics as a source of HR content; Unfortunately, these mosaics are not perfect enough estimate of HR images. The mosaicing resolves SR image from utilising overlap of video frames, and therefore it suffers from miss-registration artefacts, and not a uniform overlay of the frames, which cause nonuniform contribution of SR-resolving capabilities of the mosaicing on the entire surface of the SR image. The mosaicing is also time-consuming, making it not applicable to the real-time workflow of pCLE.
\par
In this work, similarly to~\cite{ravi2017deep}, we used a triangulation based reconstruction algorithm to simulate synthetic HR and LR endomicroscopy. However, in contrast, we took advantage of the availability of histopathological images as a source of HR signals instead of using imperfect mosaics.
During the diagnostic process, histopathological images play a similar role to pCLE. Since histopathological images are acquired with a digital camera, histology does not suffer from the problems created by irregularly distributed fibre signals.
\par
Synthetic images were created from high quality, large histopathological images. We created three sets: a training set built with 540 files, a validation set built with 227 files randomly selected from publicly available databases\footnote{\url{https://zenodo.org/record/1214456\#.XbBaDnVKhy0}},\footnote{\url{http://www.andrewjanowczyk.com/deep-learning/}},\footnote{\url{https://dataverse.harvard.edu/dataset.xhtml?persistentId=doi:10.7910/DVN/SI32FV}} of histological images from data sets~\cite{data2016,patho,DVN}, and a synthetic test set.
The synthetic test set was created with ten histopathologies from publicly available data called Kather\footnote{\url{https://zenodo.org/record/53169\#.W6HwwP4zbOQ}} published by~\cite{kather2016multi}. The synthetic test set facilitates making a comparative study between baseline solution, our proposed methodology, and classical CNN solutions. 

%% file: 4data_simulation.tex
\begin{figure}[!ht]
  \centering
  \includegraphics[width=0.85\textwidth]{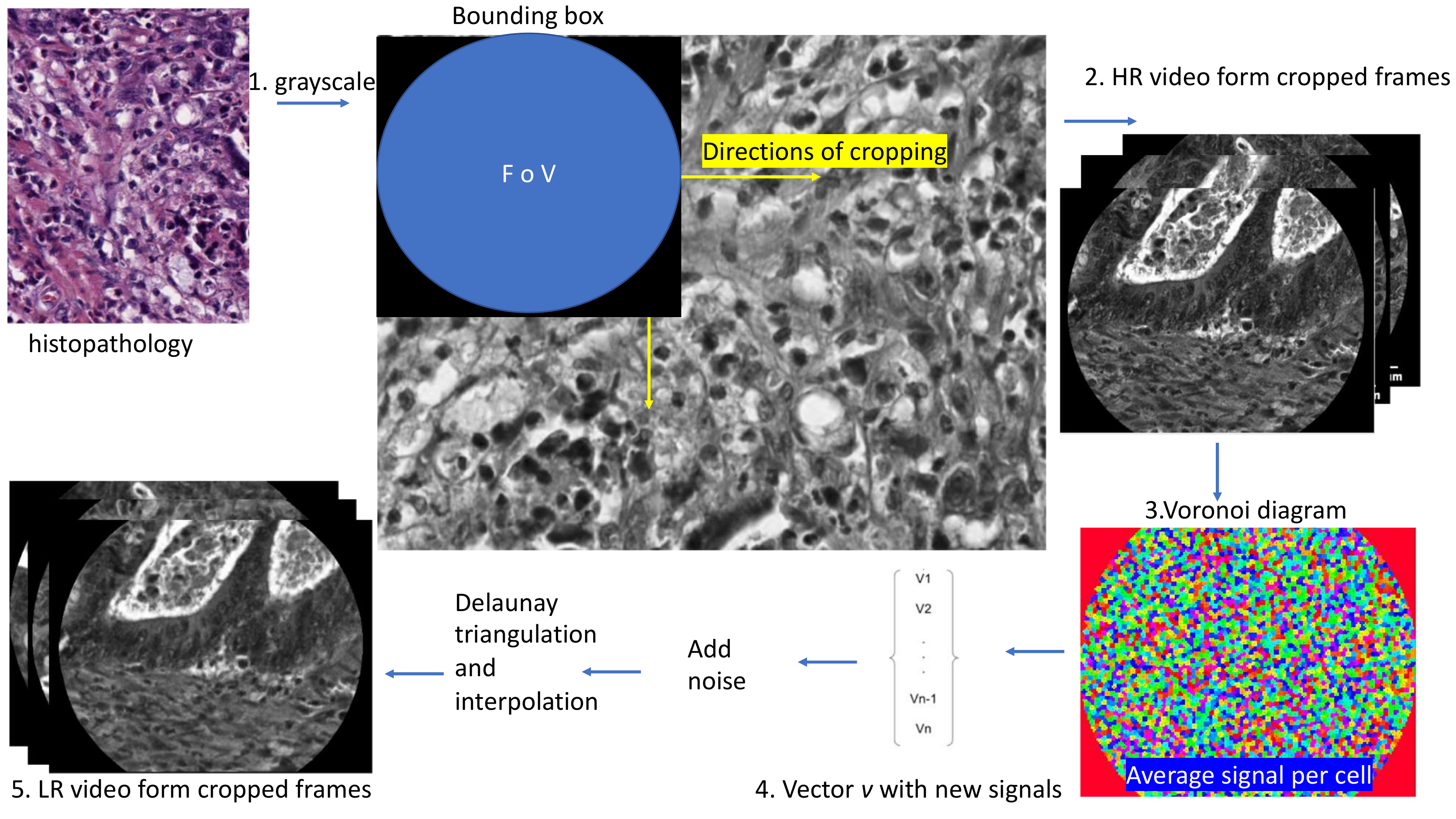}
  \caption{The illustration of the simulation for the creation of synthetic data. Histological images are transformed into synthetic endomicroscopy.}

  \label{fig:pipe}
  %\vspace{+5pt}
\end{figure}

%Ground truth images are not available in the context of pCLE, so we simulated pseudo pCLE images. 
The simulation steps are illustrated in Figure~\ref{fig:pipe}.
\subsection{Simulation of synthetic pCLE images}
%Ground truth images are not available in the context of pCLE, so we simulated pseudo pCLE images. 

\paragraph{\textbf{Synthetic HR pCLE:}} In the first part of the simulation, we transformed RGB HR histological images into grayscale HR pCLE-like videos. The simulation starts with transforming RGB images into grayscale images. Next, we randomly selected original pCLE videos from Smart Atlas~\cite{andre2011smart}, retrieved information on a bundle FoV and fibres locations for each video, and we matched pCLE metadata randomly with the histological images.
To crop pCLE-like frames from the grayscale image we were moving a bounding box of the fibre's FoV from left to right, and from top to bottom in the image, with step size equal to half of the bounding box size. These pCLE-like frames were stacked to create the synthetic HR pCLE video sequence.
\paragraph{\textbf{Synthetic LR pCLE:}}
To simulate LR pCLE videos, we used the physically-inspired pCLE-specific downsampling presented in~\cite{ravi2017deep}. In our case, the sources of irregular signals for physically-inspired pCLE-specific downsampling are HR synthetic pCLE videos. They are rich in high frequencies and pixel-level details.
\par
For every synthetic HR pCLE, we used fibres location to build Voronoi diagrams with each fibre in the centre of the Voronoi cell. Every cell corresponds to the one fibre signal, yet cell space covers several pixels around the fibre on HR image. Thus to simulate signal loss, all HR pixels in that cell are averaged, and the average HR signal is used as a new fibre signal on the LR image.
\par
Typically, pCLE noise is interpolated onto the reconstructed image from noisy signals. To achieve that, we simulate pCLE noise by adding it to the new fibre signal, before the interpolation step. We add multiplicative and additive Gaussian noise to mimic a calibration imperfection and an acquisition noise ,respectively.
\par
HR pixels with added noise are used as the irregularly sampled signals and  reconstructed to the synthetic noisy Cartesian LR image. The reconstruction is performed as interpolation based on the Delaunay triangulation~\cite{LeGoualher2004}. The synthetic LR pCLE has
the size of the HR image, yet it is characterised by the lower image quality, noise, and reduced content of information due to simulated signal loss.  
\par
Thanks to simulating signal distribution through the geometrical position of the fibres in the bundle, we simulate synthetic endomicroscopy as similar to real pCLE characterised by typical triangulation artefact, and noise patterns. In our experiments, synthetic endomicroscopies are used as a synthetic equivalent of real pCLE images. 

%First, the HR images were transformed to LR Cartesian images by physically-inspired pCLE-specific downsampling and noise addition as described in~\cite{ravi2018effective}. The only difference is that instead of using mosaics as source of HR images like in~\cite{ravi2018effective}, we use histological images. To simulate pCLE signal loss, we used fibre bundle arrangements selected from the Smart Atlas acquired with Cellvizio bundles.  

%% file: 5NW.tex
\subsection{Trainable NW kernel regression}
\label{NW_section}
\paragraph{\textbf{Input images:}} Irregularly sampled data can be represented, with an arbitrary approximation quality, on a fine Cartesian grid as the sparse artificial Cartesian image with all non-informative pixels set to zeros. To reconstruct sparse images to Cartesian images, the missing information is typically interpolated. This also means that the reconstructed images are over-sampled, and only a subset of the pixels carry information~\cite{ravi2018effective}. %Specifically, in the context of pCLE reconstruction, every 7 pixel on Cartesian images comes from interpolation of 3 neighbouring fibres. 
%As is illustrated in Figure~\ref{fig:NW}, we typically represent the corresponding sparse images by assigning a value of 0 to all non-informative pixels.
%Since the position $u,v$ of informative pixels within the sparse image $S$ is given, 

\begin{figure}[ht!]
  \centering
 \includegraphics[scale=0.48]{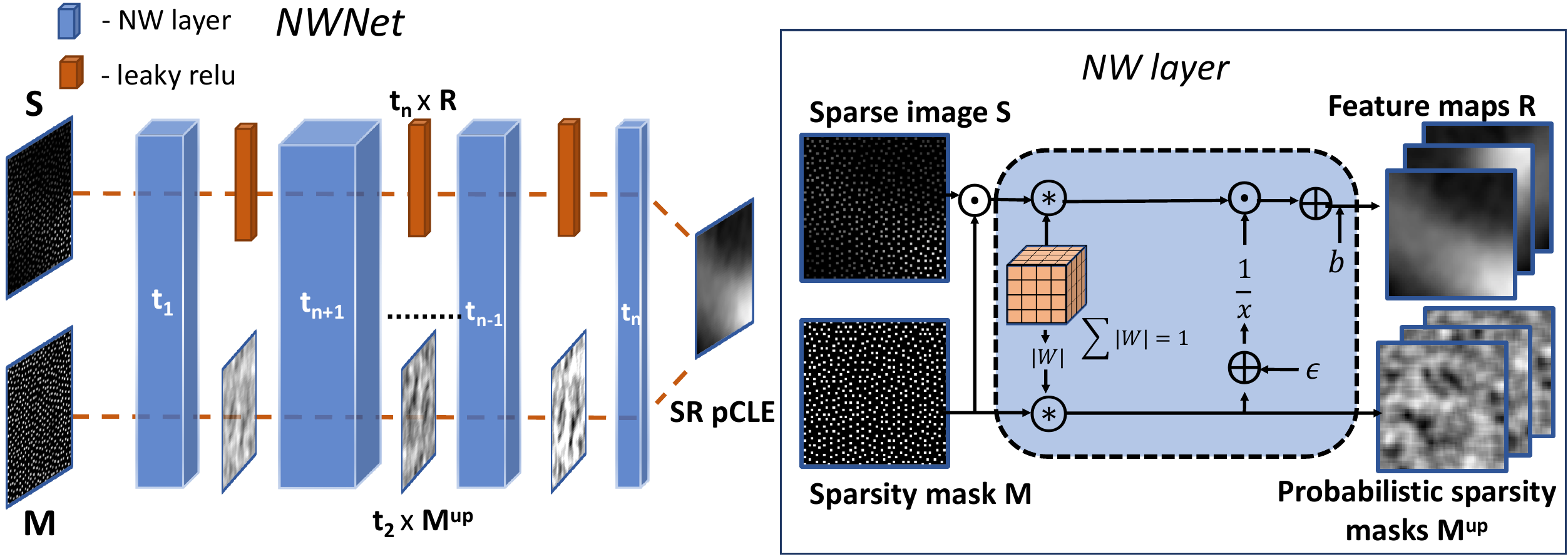}
  \caption{Graphical representation of NWNet framework (left) and NW layer (right).}%performs two convolutions  of $S$ and $M$ with shared normalised $W$; reconstructed $R$ and updated $M$. NWNet framework: the NWNet input are the sparse image $S$ and binary sparsity mask $M$; a NWNet is a stack of $n$ NW layers which utilise $M$ to learn the sparsity of $S$; the NWNet output are reconstructed feature maps $R$. Last layer fuses $n-1$ $txR$}
  \label{fig:NW}
\end{figure}

We represent the sparse pCLE image $S$ as an artificial Cartesian image. Intuitively, the image sparsity is encoded by ones and zeros for the informative and non-informative pixels respectively as a binary mask $M$ whose shape is the same as $S$. A sample $S$ and its corresponding $M$ is depicted in Figure~\ref{fig:NW}.
%in the same space. Let $M$ denote a binary sparsity mask (of the same size as image S) that encodes information such that it takes the values 1 and 0 for the informative and non-informative pixels respectively.
\par
\label{NW_layer}
% $u,v$. 
%The  output of the convolutional layer $f(X)$ is generated by convolving the input image $X$ using weights $w$ and adding a bias $b$. Weights $w$ are defined by a kernel $W$ of size $2k + 1$ along each image dimension.
%Since the irregularly distributed pCLE signal is represented on a Cartesian grid as a 
$S$ can be input into any CNN. Yet, a convolutional layer % denoted by $f_{u,v}^{}(X)= \sum_{i,j=-k}^{k} {x}_{u+i,v+j} w_{i,j}+b$,
%and it takes image $X$ as an input. The image $X$ is represented on a Cartesian grid with pixel coordinates given by $u,v$.
is defined on the Cartesian grid and considers all image pixels as equally important regardless of their position. Thus, the CNN network has to learn not only the mutual relations of informative pixels but also how to handle their sparsity, which makes the optimisation a difficult task.
 
%Several works exploring depth upsampling for sparse LiDAR data with the use of the improved constrained convolutional layers were proposed\cite{eldesokey2018propagating,hua2018normalized,uhrig2017sparsity}. These research is the most closely related to our proposed pCLE super-resolution from sparse data with NW layer. Due to the difference in the task objective and data type, it is not possible to fully benchmark networks from~\cite{eldesokey2018propagating,hua2018normalized,uhrig2017sparsity}, and major adaptation is required. 

%Here the challenge is to adapt the CNN to work around the image sparsity by predicting the missing information. To tackle this problem,
\paragraph{\textbf{NW layer:}}We compare the applicability of a classical CNN network and the proposed generalisation of NW kernel regression to pCLE image reconstruction from sparse data. 
To incorporate NW kernel regression into the CNN framework, we propose a novel trainable CNN layer henceforth referred to as  an \say{NW layer}, which models the relation of the data points by use of custom trainable kernels to perform the local interpolation. We define the core NW operation as:
% This regression technique can be efficiently implemented using two convolutions and a pixel-wise division,
%and it was successfully used with a single hand-crafted kernel to reconstruct pCLE images and mosaics~\cite{vercauteren2006robust}.
\begin{align}
     R_{u,v}^{}(S,M)=
\frac{\sum_{i,j=-k}^{k} {S}_{u+i,v+j} w_{i,j} }
{\sum_{i,j=-k}^{k} M_{u+i,v+j} | w_{i,j} |}+b
\label{eq:NW1}
\end{align}
\begin{align}
M^{up}=\sum_{i,j=-k}^{k} M_{u+i,v+j} | w_{i,j} |
\label{eq:NW2}
\end{align}
The NW layer takes as input $S$ and the corresponding $M$.
%which we generalise to allow for negative kernel values as described below. 
The mask $M$ can be seen as a probabilistic sparsity map. Initially $M\in\{0,1\}$, and the next $M$ is updated as described in equation~\ref{eq:NW2} and becomes an approximation of the probability of obtaining reconstructions $R_{u,v}$ given $S_{u,v}$. The $M^{up}$ holds the arbitrary probabilistic sparsity patterns, which are then propagated deeper to the consecutive NW layer. The outputs of the NW layer are reconstructed feature maps $R$ estimated using an NW regression and updated probabilistic sparsity masks $M$. Finally, bias $b$ is added to $R$.
The graphical representation of the NW layer is presented in Figure~\ref{fig:NW}.
\par
Classical NW kernel regression uses handcrafted, positive kernels. For a generalisation of the kernel regression, however, our trainable NW layer allows for negative values.
It is necessary for the convolution of the mask $M$ rely on the absolute value of $W$, as this operation is meant to capture the geometric influence of neighbouring pixels on the predicted values of $R_{u,v}$. 
%\input{algorithm}
%Our proposed NW layer implementation is given as pseudo-code in Algorithm ~\ref{al:NWimplementation}. 
For numerical stability, we also normalise the kernels such that $\sum_{i,j=-k}^{k} |w_{i,j}|=1$, where $w$ is a weight in position $i,j$ in the  $W$ .
\paragraph{\textbf{NWnet framework:}}
\label{NW_net}
Multiple NW layers can be stacked to generalise and benefit NW kernel regression for a irregularly sampled pCLE data.
%Given that the deeper classical CNN architecture has better performance, intuitively the same rule should apply to NWNet architectures: the deeper the NWNet, the better the generalisation. We assume that NWNet learns the sparsity of the input data, such that after a few NW layers the output features map can be directed to classical CNN layers.
This in turn facilitates end-to-end pipelines that can incorporate sparse inputs by combining NW layers with classical CNNs. As illustrated in Figure~\ref{fig:NW}, we show how to combine NW layers into deep(er) network. Each NW layer has $t$ unique kernels $W$. The first $(n=1)$ layer takes as input $S$ and binary $M$ and returns $t$ feature maps $R$ and $t$ updated sparsity masks $M_{up}$ which become the input for the next NW layer. The last NW layer of the NWNet framework returns only $t$ feature maps $R$, and masks $M$ are discarded.
\paragraph{\textbf{Application to endomicroscopy image reconstruction:}}
\label{application}
NWNet in combination with complex deep learning models, such as EDSR~\cite{lim2017enhanced}, may allow for reconstruction of higher quality pCLE than the more typically used interpolation.
%It was shown in \cite{ravi2018effective} that Enhanced Deep Residual Networks for Single Image Super-Resolution (EDSR)~\cite{lim2017enhanced} allows for improving the quality of over-sampled Cartesian pCLE images. 

\begin{figure}[ht]
  \centering
  \includegraphics[scale=0.35]{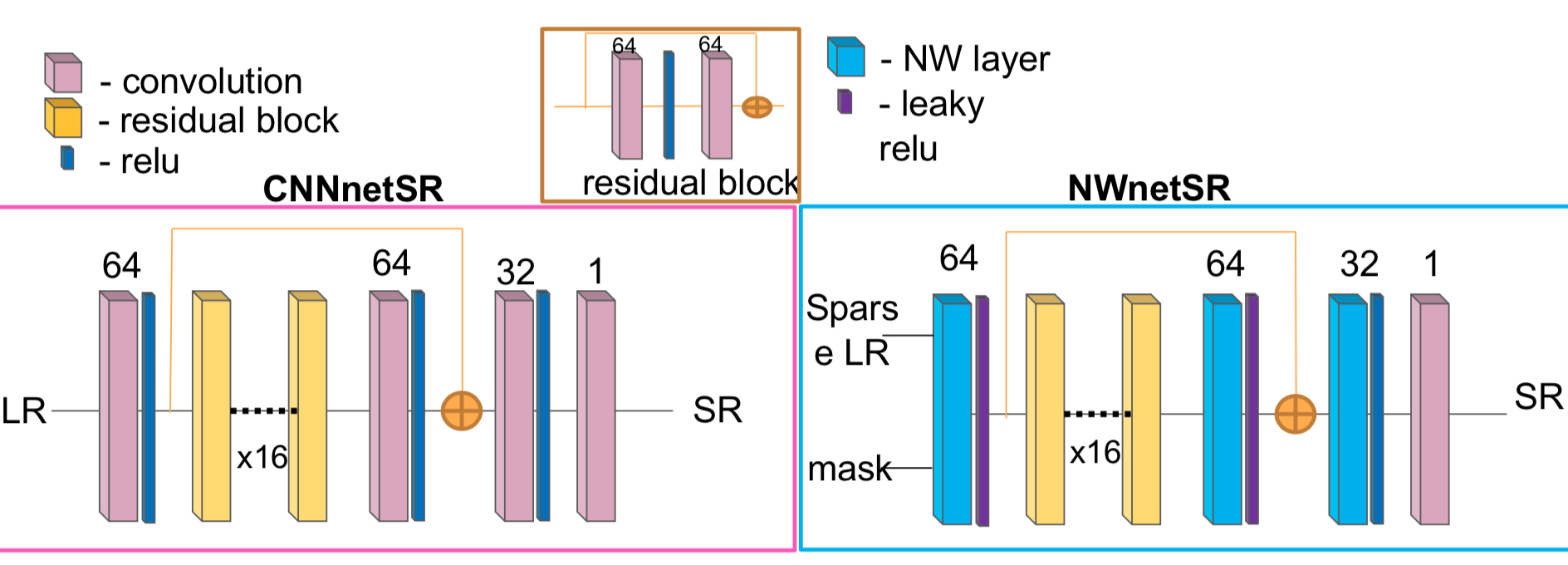}
  \caption{The graphical representation of architectures CNNnetSR and NWnetSR.}
  \label{fig:edsr}
  %\vspace{+5pt}
\end{figure}
\par
Inspired by EDSR network~\cite{lim2017enhanced}, we design two architectures CNNnetSR and NWnetSR presented in Figure~\ref{fig:edsr}. 
CNNnetSR was designed to perform the SR task from Cartesian or sparse LR pCLE images. The network is very similar to EDSR architecture~\cite{lim2017enhanced}, but with two small improvements.
First, we do not use upsampling layer, because synthetic LR and HR have the same size. In place of upsampling layer, we put the convolution layer with 32 filters. Second, the last convolutional layer aims at fusing the output feature maps from the penultimate layer into the Cartesian image. We find it more beneficial to use the kernel of size one, which is commonly used to reduce the number of features maps, than three. All convolutional layers have the kernel size 3. The Last layer uses a linear activation function. 
\par
We want to take advantage form SISR for NW kernel regression, so we design NWnetSR based on CNNnetSR by partly replacing standard convolution with NW layers. For fist NW layer, the kernel size is 9 across each image dimension. The size was chosen based on the known distribution of fibres across a Cartesian image to ensure that each convolution would capture more than 10 informative pixels. For deeper NW layers kernel size is 3.
The NW weights were initialised with a truncated normal distribution with mean, and standard deviation equal to 0.2 and 0.05 respectively. 

%% file: 6nw_implementation.tex
\subsection{Implementation details}
\label{experiments}
%Additionally, we compared our results to EBSR, which takes as input reconstructed pCLE images as presented in~\cite{ravi2018effective}. 
%The synthetic pCLE dataset published in~\cite{ravi2018effective} was generated with a registration-based simulation which produces synthetic high-resolution (HR) estimates of ground truth ime used 
\paragraph{\textbf{Data pre-processing}}
To facilitate training, video sequences were normalised for each frame individually by subtracting mean and standard deviation of LR frame as follows: $LR = (LR - mean_{LR})/std_{LR}$ and $HR =( HR - mean_{LR})/std_{LR}$, and then scaled to the range [0,1]. Synthetic LR images were transformed to the synthetic sparse LR images by setting zeros to all pixels which do not correspond to any fibre signal. 
The masks were generated as a binary image, where fibre positions from the LR image are set to ones. Lastly, to perform batch-based training, we extracted non-overlapping $64 \times 64$ sparse and Cartesian patches for the train and validation sets. The test sets were built with sparse and Cartesian full-size synthetic pCLE images. 
\paragraph{\textbf{Training strategy}}
To achieve the best training results for each model individually, networks were trained with Population Based Training (PBT)~\cite{jaderberg2017population}. 
PBT is it an optimisation technique design to find the best training parameters for the network. During PBT training, a population of the models with different parameters is trained, these models are periodically validated, and the weights from the best performing model in the population are copied to other members of the population.
We set the population size to 6 workers, where each member of the population uses a single GPU optimising a network for epoch in one PBT iteration for a total of 100 iterations. The perturbation interval was set for every 20 iterations. The hyperparameter search applied to the 6 learning rates, which were initially set to $10 ^{i}$ where $i\in\{-2,-3,...,-7\}$.
We used the Adam optimiser and set $\beta_{1}$=0.9, $\beta_{2}$=0.999, $\epsilon=10^{-8}$.
Based on results presented in \cite{ravi2018effective}, the models were trained with SSIM+L1 loss~\cite{zhao2015loss}. Finally, the best performing model from the population is used to generate results on the test sets described in Section~\ref{data_sim}.

%% file: 7results_validation.tex
\label{results}
We compared performance of three DL models: CART for CNNnetSR trained with  Cartesian input, SPARSE for CNNnetSR trained with sparse input and NW
for NWnetSR trained with sparse input and corresponding mask.
These models were handling irregular signals as input differently. As a baseline method, we used the golden standard reconstructions algorithm currently implemented in the clinical setup, which is based on linear interpolation and Delaunay triangulation~\cite{vercauteren2006robust}. We also provide a comparison to reconstructions obtained using NW kernel regression with a single Gaussian kernel~\cite{tom_phd}. The final performance of the models is evaluated by comparing reconstructed SR pCLE with the HR synthetic images from the test set.
\par
To quantify how the standard convolution performs on sparse pCLE images in comparison to using Cartesian reconstructions as the input, we trained two unique models based on CNNnetSR network: CART trained using reconstructed Cartesian images, and SPARSE one trained with sparse images.
\par
To test NW kernel regression benefits from generalisation via learning multiple kernels, we trained the NWnetSR as SISR network for the task of pCLE SR reconstruction with sparse input images and masks.
\par
To measure the image quality of the SR pCLE reconstructions, we design an image quality assessment (IQA) procedure which consists of two complementary metrics typically used for this task:
peak signal-to-noise ratio (PSNR) and the Structural SIMilarity index  SSIM~\cite{wang2004image}. 
\par
\begin{figure}[!ht]
  \centering
  \includegraphics[scale=0.49]{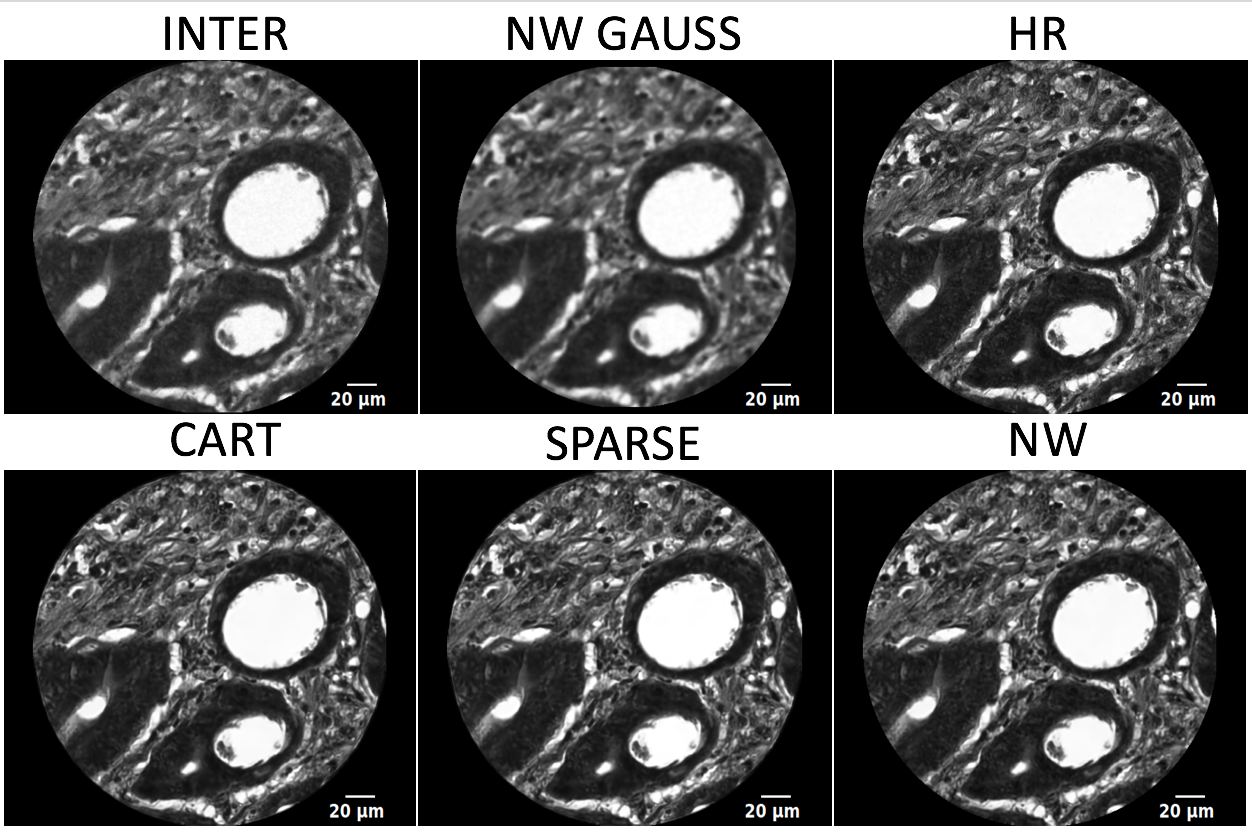}
  \caption{Sample reconstructions from 6 methods handling sparse data in different ways.}
  \label{fig:images}
  %\vspace{+5pt}
\end{figure}
The results computed for the images from the test set, described in Section~\ref{data_sim}, are shown in Table~\ref{tab:IQA}. The main observation is that each of DL model, including NW outperforms the baseline interpolation technique. Training of the NWnetSR as SISR network, generalise NW kernel regression and outperforms traditional NW regression with hand-crafted Gaussian kernel, proving that use of many kernels which are estimated on the data set are more beneficial then custom single Gaussian kernel.  %Models trained on synthetic data translate to the pCLE domain.
There is no significant improvement for PSNR and SSIM scores when comparing DL models between each-other. They perform almost indistinguishable.   
We also provide example reconstructions in Figure~\ref{fig:images}, and it can be noticed easily that all DL images are almost identical.
\input{table_syntetic}

We analysed SR reconstructions qualitatively, and we can observe two tendencies for all DL models. First, SR reconstructions differ slightly on a pixel level, but that differences do not affect how the video is perceived as a whole, and maybe a reason for slightly different metrics score during IQA. Second, in the opinion of our clinical collaborators, SR reconstructions are more aesthetically pleasing than baseline reconstructions INTER and NW GAUSS. Neither of SR has triangulation artefacts, and every SR reconstruction has significantly reduced noise, additionally benefiting from improved contrast and visibility of details.\par
The results confirm that the NW layer among with other layers handling sparse data is a choice for image reconstruction and yield increasingly good image quality results from sparse pCLE data to Cartesian SR image.

%% file: table_syntetic.tex
\begin{table}[htbp]
  \centering
  \caption{IQA: comparison on average performance of 5 methods aimed at pCLE image reconstructions. Table shows average PSNR  and SSIM score with standard deviation for all frames in 10 pseudo pCLE videos.}
\begin{tabular}{|c|c|c|}
\hline
model name & SSIM & PSNR  \\
\hline
NW GAUSS &0.758$\pm$0.025 &25.88$\pm$1.0 \\
\hline
INTER & 0.800$\pm$0.018 & 24.65$\pm$1.2\\
\hline
CART  & 0.902$\pm$0.015 &30.98$\pm$1.1 \\
\hline
SPARSE & 0.907$\pm$0.016 & 30.99$\pm$1.0 \\
\hline
NWNET &0.900$\pm$0.015 &30.96$\pm$1.1\\
\hline

\end{tabular}
\label{tab:IQA}%
\end{table}%

%% file: 8conclusions.tex
\label{conclusions}
%This work advances deep learning research by introducing a novel NW layer.
The proposed CNN layer enables the use of sparse images as input to the CNNs and learns the sparse image representation.
In the context of pCLE, this is the first work which proposes end-to-end deep learning image reconstructions from irregularly sampled fibre data.\par
%prior information to super-resolved pCLE images . 
NW layer is used as a building block in dedicated architecture NWNetSR which performs deep generalised Nadaraya–Watson kernel regression. NWNetSR capture data sparsity and learn reconstruction kernels for sparse data.
We demonstrated that among with other CNN-based solutions, NWNetSR also improves the image quality of pCLE images.
We proved that super-resolved pCLE images have better quality than the one interpolated with use of the baseline method; thus the proposed super-resolution pipeline outperforms the currently used reconstruction method. \par
%Moreover, shallow NWNets outperform EDSR for Cartesian images. It leads to the conclusion that NWNets due to their principled handling of sparse data, generate output features maps which capture a better representation of the input sparse data than interpolated Cartesian images. 
%NW layer and its extension to NWNets give an efficient way to incorporate irregularly sampled data as the input of any CNN pipeline for regularly sampled data. 
Deep learning methods for regularly sampled data may be transferred to sparse data by adopting NW layer in standard CNN networks.
We have shown the successful implementation of the reconstruction pipeline, which combines NW and CNN layers, and is trained in a supervised manner as SISR, reconstructing super-resolved images from sparse input images.

%Beyond pCLE, we believe that our research
%may benefit other applications in which data is defined on a graph structure. NWNets are computationally efficient and can readily be adapted to either graph-in/scalar-out or graph-in/image-out. Future work will focus on developing NWNets architectures and applying NWNets to tasks such as classification. 

%% file: 9acknowladgment.tex
This work was supported by Wellcome Trust: 203145Z/16/Z; WT101957; 203148/Z/16/Z, and EPSRC: NS/A000050/1; NS/A000027/1; EP/N027078/1. This work was undertaken at UCL and UCLH, which receive a proportion of funding from the DoH NIHR UCLH BRC funding scheme. The PhD studentship of Agnieszka Barbara Szczotka is funded by Mauna Kea Technologies, Paris, France.
Tom Vercauteren is supported by a Medtronic / Royal Academy of Engineering Research Chair: RCSRF1819/7/34.

%% file: 9conflict.tex
\textbf{Conflict of Interest:} The PhD studentship of Agnieszka Barbara Szczotka is funded by Mauna Kea Technologies, Paris, France. Tom Vercauteren owns stock in Mauna Kea Technologies, Paris, France.
The other authors declare no conflict of interest.\\
\textbf{Ethical approval:}  All procedures performed in
studies involving human participants were in accordance with the ethical standards of the
institutional and/or national research committee and with the 1964 Helsinki declaration and its
later amendments or comparable ethical standards.
\\
\textbf{Informed consent:} For this type of study formal consent is not required. This article does not contain patient data.